# Automatic exposure volumetric additive manufacturing

Antony Orth[1]*, Yujie Zhang[1], Katherine Houlahan[1], Daniel Webber[1], Hao Li[1], Nicolas Milliken[1], Joshua Latimer[2], Derek Aranguren van Egmond[1], Bosco Yu[2], Jonathan Boisvert[1], Chantal Paquet[1]

[1]National Research Council Canada, Ottawa, Canada

[2]University of Victoria, Victoria, Canada

*antony.orth@nrc-cnrc.gc.ca

## Abstract

Tomographic volumetric additive manufacturing (VAM) achieves high print speed and design freedom by continuous volumetric light patterning.  This differs from traditional vat photopolymerization techniques that use brief sequential (2D) plane- or (1D) point-localized exposures.  The drawback to volumetric light patterning is the small exposure window.  Overexposure quickly leads to cured out-of-part voxels due to the nonzero background dose arising from light projection through the build volume.  For tomographic VAM, correct exposure time is critical to achieving high repeatability, however, we find that correct exposure time varies by nearly 40% depending on resin history.  Currently, tomographic VAM exposure is timed based on subjective human determination of print completion, which is tedious and yields poor repeatability. Here, we implement a robust auto exposure routine for tomographic VAM using real-time processing of light scattering data, yielding accurate and repeatable prints without human intervention.  The resulting print fidelity and repeatability approaches, and in some cases, exceeds that of commercial resin 3D printers.  We show that auto exposure VAM generalizes well to a wide variety of print geometries with small positive and negative features.  The repeatability and accuracy of auto exposure VAM allows for building multi-part objects, fulfilling a major requirement of additive manufacturing technologies.

## Introduction

Tomographic volumetric additive manufacturing (VAM) has recently emerged as a flexible and ultra-rapid approach to polymer 3D printing[1,2].  Like other vat photopolymerization techniques, tomographic VAM works by patterning near-UV light inside a photocurable resin.  When the local absorbed light dose reaches a critical value, the liquid resin cures, resulting in solid polymer.  The shape of the solid polymer follows the shape of the patterned light dose inside the printing volume, enabling printing of solid 3D objects.  Tomographic VAM differs from traditional polymer 3D printing methods, such as digital light processing (DLP) or stereolithography (SLA), in the way that the light is patterned[3].  In DLP, individual 2D images - each corresponding to a layer of the desired object - are projected into the photoresin sequentially.  In between each 2D image exposure, the build platform is moved vertically, allowing for the next layer to be added on to the bottom of the print.  SLA works in much the same way, but instead a laser beam is traced over a 1D trajectory that approximates the shape of a layer.  These serial approaches work well but are inherently slow because only a small amount of resin is illuminated at any given time.  Furthermore, the inter-layer mechanical movements add significantly to the total print time.  Continuous fabrication projection lithography techniques such as continuous liquid interface production (CLIP) and high area rapid printing (HARP) avoid discrete layer steps, but are still limited in speed by resin replenishment under the build plate, especially for viscous resins[4,5].

In tomographic VAM, the whole build volume is illuminated for the entire print process.  This is achieved by projecting a sequence of images through a rotating cylindrical vial of photocurable resin.  The projected images are calculated such that the total absorbed light dose inside the rotating photoresin volume approximates the shape of the desired object[1,2].  This entire-volume exposure speeds up printing and avoids the need to support overhangs with scaffolding material, enabling more complex prints and smooth surfaces free of support structure blemishes and layer step artifacts [6]. Since its initial demonstration[1,2], significant progress has been made to tomographic VAM print fidelity[7–9], projection algorithms[10–12], speed[13], material palette[14–22] and ease of use[23,24].

Despite this progress, the topic of print repeatability has been entirely absent from the tomographic VAM literature. Readers are generally given no indication as to how many prints were needed to obtain the handful of examples showcased in each paper. Moreover, on the rare occasion when print *accuracy* is measured, it is not compared with commercial vat photopolymerization printers[1,7,8]. This is likely because, to-date, tomographic VAM has required manual exposure control, resulting in large variation in print quality between users. Exposure times are also highly dependent on resin preparation and age, which can currently only be compensated for by human input. This is a serious impediment to wider adoption of tomographic VAM, because effective use of a tomographic VAM printer currently requires significant user skill. Despite its promise as a fast 3D printing technique, tomographic VAM is, ironically, very time intensive because of the need for manual intervention on every print to ensure good exposure. Ideally, tomographic VAM would operate more like a commercial DLP or SLA printer, where the correct optical exposure is determined without user intervention, resulting in "set it and forget it" operation.

In this work we present the first VAM system with automatic exposure (AE, AE-VAM), eliminating user intervention during printing. This is achieved using an imaging system to monitor the total scattered light signal during the print[8]. We previously used this imaging system to enable real time reconstruction of the printed object. In this work, we use the total scattering signal along with a novel calibration procedure to identify a geometry-independent exposure endpoint for each print. We test our approach on a wide range of geometries using an acrylate resin optimized for VAM. For the commonly used 3DBenchy test print[25], AE-VAM achieves an average root-mean-squared (RMS) print surface error of 0.100mm (accuracy), and inter-print surface variation of 0.053mm (repeatability) across 25 VAM-printed 3DBenchy models (largest dimension 14.38 mm). Crucially, this is achieved despite a print time variation of up to 39%, which occurs when re-using resin. The ability to compensate for variation in print timing due to resin re-use is a unique need for VAM, where the entire vat of resin is exposed in each print. Without inline adjustment of print time, correct exposure would only be possible with fresh resin, resulting in a significant amount of waste and increased cost. Instead, our approach enables resin recycling, with no significant degradation in print quality for at least 4x re-use. This approach results in commercial-grade repeatability and state-of-the-art print fidelity, accessible to users without any prior experience using a VAM printer.

**Experimental**

Our AE-VAM system is shown in Figure 1a. The printer consists of a near-UV projector (EKB 4500-MKII, 405nm, 7.6 μm pixel pitch) placed in front of a glass vial filled with photoresin. The vial is mounted on a continuous rotation stage (Zaber X-RSW-E) which is set to rotate at 40 deg/s during printing. While printing, the projector illuminates the rotating resin volume with a sequence of pre-calculated images at a rate of 1 image/degree. These images are calculated such that after an integer number of vial rotations, the absorbed light dose profile in the resin matches the 3D shape of the desired object. Details of the calculation process can be found in the Methods section.

The key component for automatic exposure is the imaging system, which was first described as Optical Scattering Tomography[8]. This illumination component consists of a quasi-collimated red light emitting diode (LED, Thorlabs M660L4, 660nm) oriented along the axis of the vial, and a camera (FLIR GS3-U3-23S6M ) with lens that observes the vial, with an optical axis oriented perpendicular to the vial axis. The build volume is illuminated approximately uniformly by the red LED. The camera is also fitted with a bandpass filter (Thorlabs FBH660-10) that matches the illumination wavelength of the LED. The liquid resin has a uniform refractive index and therefore does not scatter light appreciably, resulting in a dark image on the camera. Once the magnitude of light dose absorbed by the resin reaches a threshold, the curing reaction starts. At this point, solid microscopic regions of solid polymer, with a higher refractive index than the liquid monomer resin, begin to form. This results in significant scattering, which increases as the reaction proceeds. We have shown previously that this enables 3D mapping of the evolving print geometry during the curing process. This requires computational backprojection of the recorded images, which can be done in real time. However, in this current work, we leverage these live 2D images in a simpler way to identify the correct print exposure.

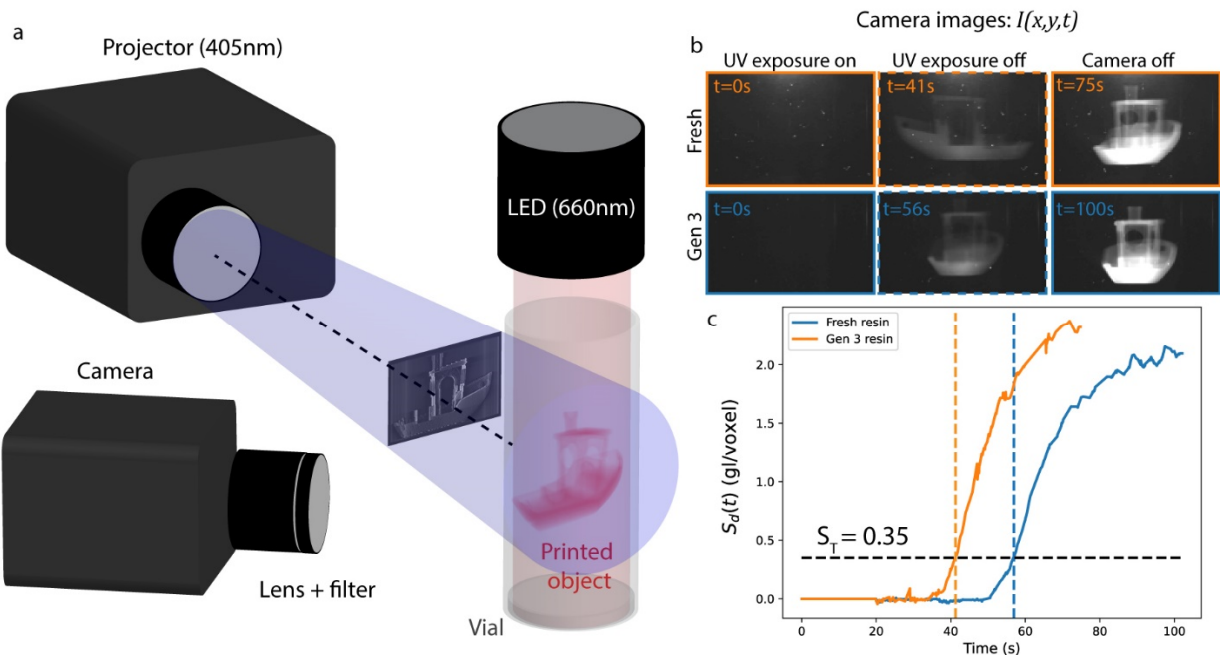

Fig. 1. a) Schematic of the VAM printer. A projector projects a sequence of images through a rotating vial of photocurable vial. An overhead red LED illuminates the resin; scattered red light is recorded by a camera. b) Example images recorded by the camera during printing. Image brightness was scaled up by a factor of 1.25 for display purposes. c) Temporal behaviour of the scattering density $S_d(t)$ during a print for fresh resin (blue curve), and Gen 3 resin (orange curve). Dashed vertical lines indicate the time at which the projector UV exposure is terminated. The dashed horizontal line indicated the threshold scattering density at which the projector UV exposure is terminated.

To monitor the print progression, we calculate the scattering density $S_d$ from the 2D camera image at time $t$ ($I(x,y,t)$, Fig. 1b) and the volume of the desired object $V$:

$$S_d(t) = \sum_{x,y} I(x,y,t)/V - \sum_{x,y} I(x,y,0)/V \qquad (1)$$

We represent $S_d$ with units of camera gray levels per voxel (gl/voxel). The second term on the right hand side of Equation 1 is a background subtraction to remove the effect of the small initial scattering of the liquid monomer, dust and debris and camera dark counts. $S_d(t)$ initially increases with the proliferation of small polymer chains in the liquid monomer, as shown in Fig. 1c. This phenomenon is known as critical opalescence[26], where in our case where large regions of dissimilar phases (polymer/monomer) with unequal refractive indices scatter light into the camera. We use $S_d(t)$ as a reaction state variable that reports on the progression of the curing process.

We hypothesize that if we terminate projection exposure when $S_d(t)$ reaches a fixed threshold value $S_T$ (and therefore a consistent reaction state), that we will obtain consistent prints. Termination times are shown as vertical dashed lines in Fig. 1c for two example prints. Crucially, for uniform illumination, $S_d(t)$ is independent of the orientation, location, geometry and volume of the polymerizing object, making it a robust parameter for process control.

**Calibration**

To identify the $S_T$ value that gives optimal prints, we perform a series of calibration prints (Fig. 2). The calibration object is a 12mm diameter x 2mm thick disk, with a series of half disks at the periphery (Fig. 2a). The half disks are either connected to the large disk by a two spokes or are separated from the large disk by a gap. Each half disk connected by spokes has a unique spoke width, and likewise for half disks disconnected by gaps. Spoke widths vary from 1-7 pixels (px) in steps of 1px, and the gap widths vary from 2-8px in steps of 1px. This calibration object is rasterized on a grid with pixel size 67μm, yielding spoke widths from $67 - 459$μm and gap widths from $134 - 536$μm. Spokes and gaps are not generally aligned to the rectangular pixel grid and are therefore linearly interpolated, as shown in Fig. 2a. The rasterized 2D object model is then extruded in the vertical direction to form a greyscale 3D model. For an ideal print, all the half disks with spokes will be connected to the large disk, whereas all the gapped half disks will be disconnected. In practice, overexposing the print will result in all the spokes forming, but will also connect the gapped half disks to the large disk. For under exposed prints, the gapped half disks will not connect to the large disk, but the thinner spokes may also fail to cure adequately, resulting in missing spoked half disks. The goal is to find the $S_T$ value that results in large number of connected spoked half disks, and a small number of gapped half disks connected to the main large disk. A photograph of a typical calibration print is shown in Fig. S1.

For each value of $S_T = \{0.15, 0.25, 0.35, 0.45, 0.55\}$ gl/voxel we printed 4 calibration disks, where UV projection was terminated when the $S_d(t) = S_T$. After exposure, washing and drying, each disk is visually inspected to identify which half disks remain connected to the main disk. X-ray CT (Bruker Skyscan 1275) was also used to confirm the connected half-disks, though this is not necessary. The mean values of the smallest spoke and smallest successful gap for each $S_T$ value are shown in Fig. 2b, along with their standard deviations over the N=4 replicates. We also plot a figure of merit (FOM) defined by:

$$\text{FOM} = \text{\# spoked half disks connected} - \text{\# gapped half disks connected} \qquad (2)$$

This FOM measures the tradeoff between under and overexposure, with a larger FOM being desirable. An ideal print would yield $\text{FOM} = 7$ because all 7 spoked half disks would be connected, compared to none of the gapped half disks. We find that the mean FOM > 2 for $0.15 \leq S_T \leq 0.35$. We call this the "acceptable exposure range", where the resolution and positive and negative features can be traded off against one another. For all the subsequent prints in this paper, we use $S_T = 0.35$. It is desirable to work on the high exposure edge of the acceptable exposure range because of the low degree of cure achievable in tomographic VAM due to finite dose contrast. The FOM drops to $\approx 0$ for $S_T = 0.45, 0.55$, representing significant over exposure causing over curing. For $S_T < 0.15$, the calibration disk does not form properly due to under exposure.

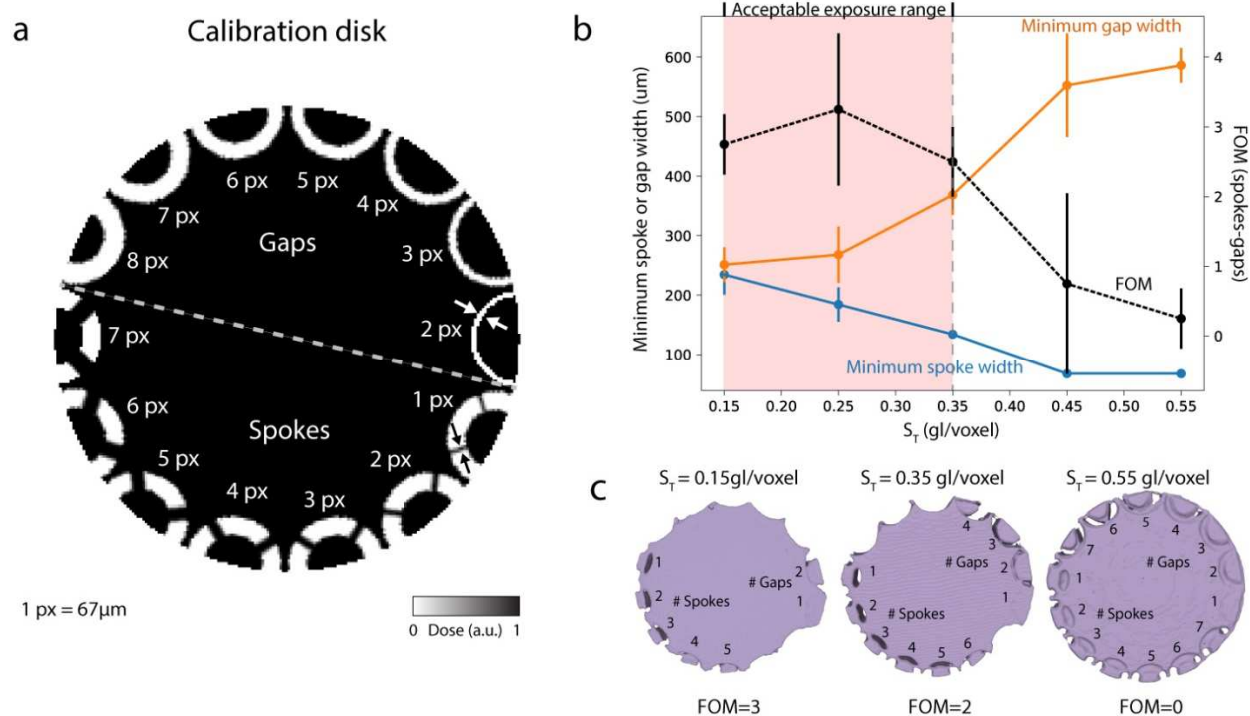

Fig. 2. a) Rasterized image of the calibration disk. The spoke and gap widths in pixels are indicated. The relative target projected dose is shown in colormap on the bottom right. b) The minimum spoke and gap widths for each $S_T$ value. The dashed black line graph (right-hand axis) shows the mean figure of merit (FOM) = # connected spoked half disks - # connected gapped half disks for each $S_T$ value. The units of $S_T$ are gray levels / voxel ($gl/voxel$). The pink region shows the acceptable exposure range, for which the FOM>2. The vertical dashed line indicates the threshold value used for the prints in this paper: $S_T = 0.35\ gl/voxel$. Error bars indicate the standard deviation over N=4 replicates for each $S_T$ value. c) X-ray CT renderings of typical calibration disks printed with $S_T = 0.15, 0.35, 0.55\ gl/voxel$. FOMs for each individual disk are indicated.

**Accuracy and Repeatability**

Having identified an appropriate $S_T$ (exposure) value, we turn our attention to quantifying the print accuracy and variation of our auto exposure approach.

To assess print fidelity for a complex geometry, we printed 25 3DBenchy models, 11.5 mm in height, with AE-VAM using an auto exposure threshold of $S_T = 0.35$. The 3DBenchy model is a benchmark in additive manufacturing and contains many challenging details including small positive and negative features[25]. The 3DBenchy model has yet to be successfully printed with VAM, with either large deformations or more subtle features failing to appear.

The reference 3DBenchy model, along with a typical AE-VAM 3DBenchy print result (as imaged using micro X-ray CT), are shown in Figs. 3a and b. Photographs of typical AE-VAM printed 3DBenchys and X-ray CT renderings of all AE-VAM printed 3DBenchys are shown in Fig. S2. For comparison, prints of the same model with commercial vat photopolymerization 3D printers are shown in Figs. 3c and d. The first 5 3DBenchy models were printed with fresh resin (generation (gen) 1). One 3DBenchy was printed in each of 5 nominally 1-inch diameter vials. A single 3DBenchy print consumed between 10 - 15% of the resin in the vial. The remaining liquid resin was stirred vigorously, and then left to sit overnight for bubbles to disappear. The next day, 5 more 3DBenchys were printed in this once re-used resin (gen 2). This process was repeated until 25 total 3DBenchys were printed, with the last 5 3DBenchys printed in 4x re-used resin (gen 5). The ability to produce accurate prints with reused liquid resin is vital for the practical use of VAM, because a large fraction of resin in the build volume is exposed but not cured in each print. This partially exposed resin has a faster curing time: 3DBenchys print time decreased from 56.8s to 39.8s from fresh to gen 5 resin (Fig. 4a). Consequently, one cannot expect to achieve repeatable printing by calibrating to a fixed exposure time. Without automatic exposure compensation, achieving consistent prints would require that the entire resin volume is discarded after every print, leading to significant waste.

For print quality quantitation, we use micro x-ray CT imaging to obtain a 3D model of the final printed, washed and post-cured parts, with voxel size 0.025mm. We use Meshlab[27] to align the micro x-ray CT meshes with the reference 3DBenchy file, and subsequently calculate the mean and RMS surface deviation of each printed 3DBenchy from the reference part. The results are summarized in Fig. 4b and Table S1. We found that the mean RMS surface deviation from the printed 3DBenchys to the reference geometry across all 25 3DBenchys is 0.100mm, corresponding to 0.7% of the largest 3DBenchy dimension (length 14.3mm), or 4 x-ray CT voxels. The mean surface deviation is $-0.013$mm. This is significantly less than the *projected* voxel size in the resin ($0.045 \times 0.045 \times 0.067$mm). Interestingly, the RMS print error (Fig. 4b) decreases slightly for re-used resin. This may be due to the faster print times, which reduces the negative impact of oxygen diffusion[9], though further investigation is required.

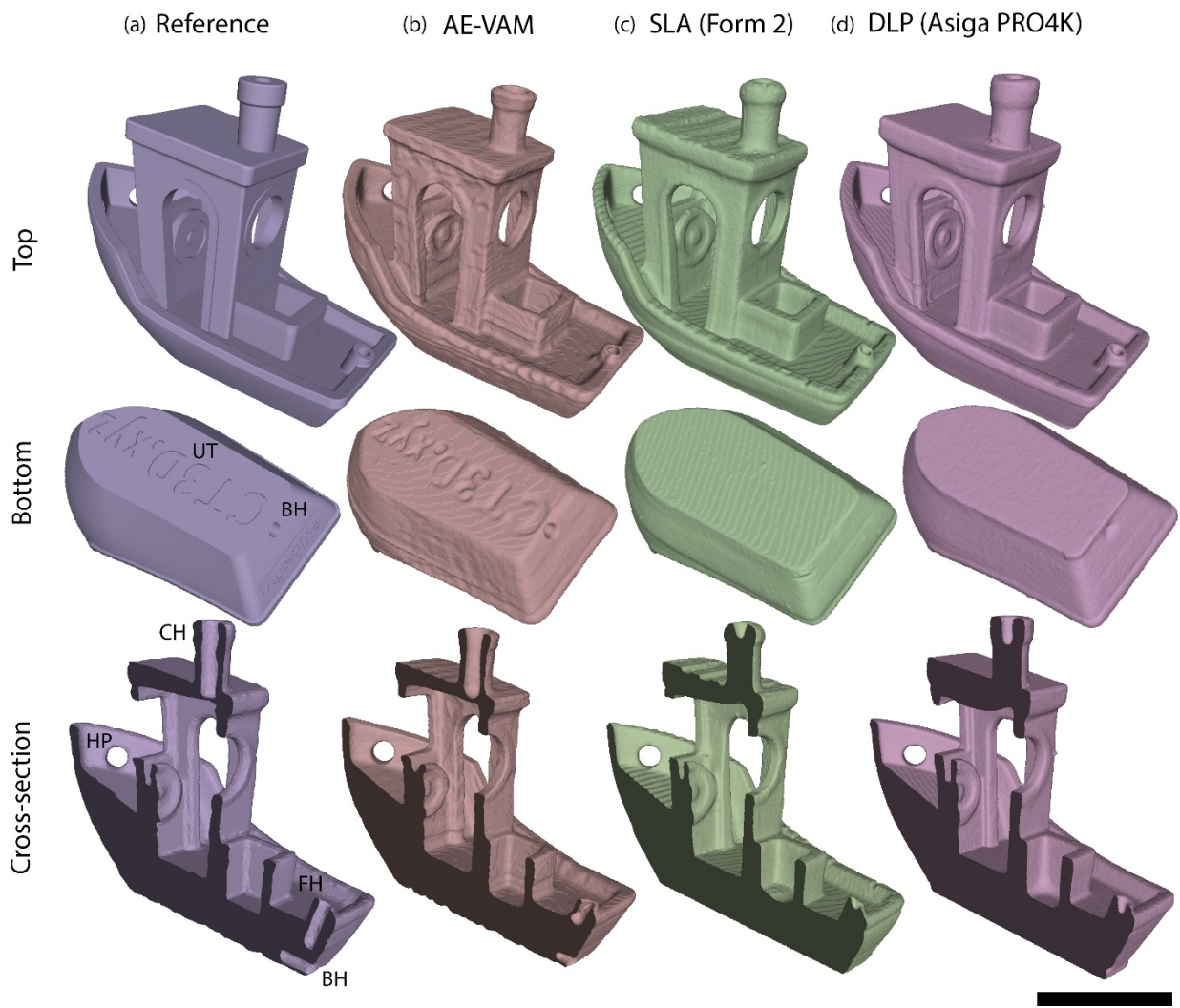

Fig. 3. Comparison of 3DBenchy models printed with VAM, SLA and DLP. a) The reference model. Subtle features such as the underside text (UT), blind hole (BH), chimney hole (CH), hawespipes (HP), and flagpole hole (FH) are indicated on the bottom and cross-sectional views. b) AE-VAM printed model. c) SLA (Formlabs Form 2) printed model. d) DLP (Asiga PRO4K) printed 3DBenchy model. Many of the small negative features (UT, BH, CH, FH) are filled in for SLA and DLP prints but appear correctly in the VAM-printed model. Scale bar is 5mm.

While the RMS surface deviation from the reference object indicates the accuracy of our AE-VAM printer, the inter-print surface deviation indicates its repeatability. To quantify repeatability, we calculate the RMS surface deviation between each pair of 3DBenchys within each generation of resin (Fig. 4c). For gen 1 and gen 2 resins, the inter-print RMS surface deviation is 0.045mm and 0.041mm, increasing to 0.069mm for generation 5 resin (Table S1). The mean inter-print RMS surface deviation is 0.053mm when averaged over all generations of resin.

In addition to AE-VAM's commercially competitive RMS deviation, we observe that all the small positive and negative features of the 3DBenchy form correctly (Fig. 3 "bottom" and "cross-section"). This includes steering wheel (SW), chimney (CH), hawsepipes (HP), underside text (UT), rear flagpole hole (FH) and horizontal blind hole (BH). These features consistently appear in our 3DBenchy prints, although the thin wall of the flagpole hole ($\approx$ 0.16mm) sometimes does not appear for gen 1 resin (likely due to oxygen diffusion[9]), or collapses during the washing step. The only feature that does not appear is the #3DBenchy nameplate at the rear of the boat, which is not rasterized from the original CAD file due to its small thickness. We believe that this is the first demonstration of a VAM-printed 3DBenchy containing all resolvable features.

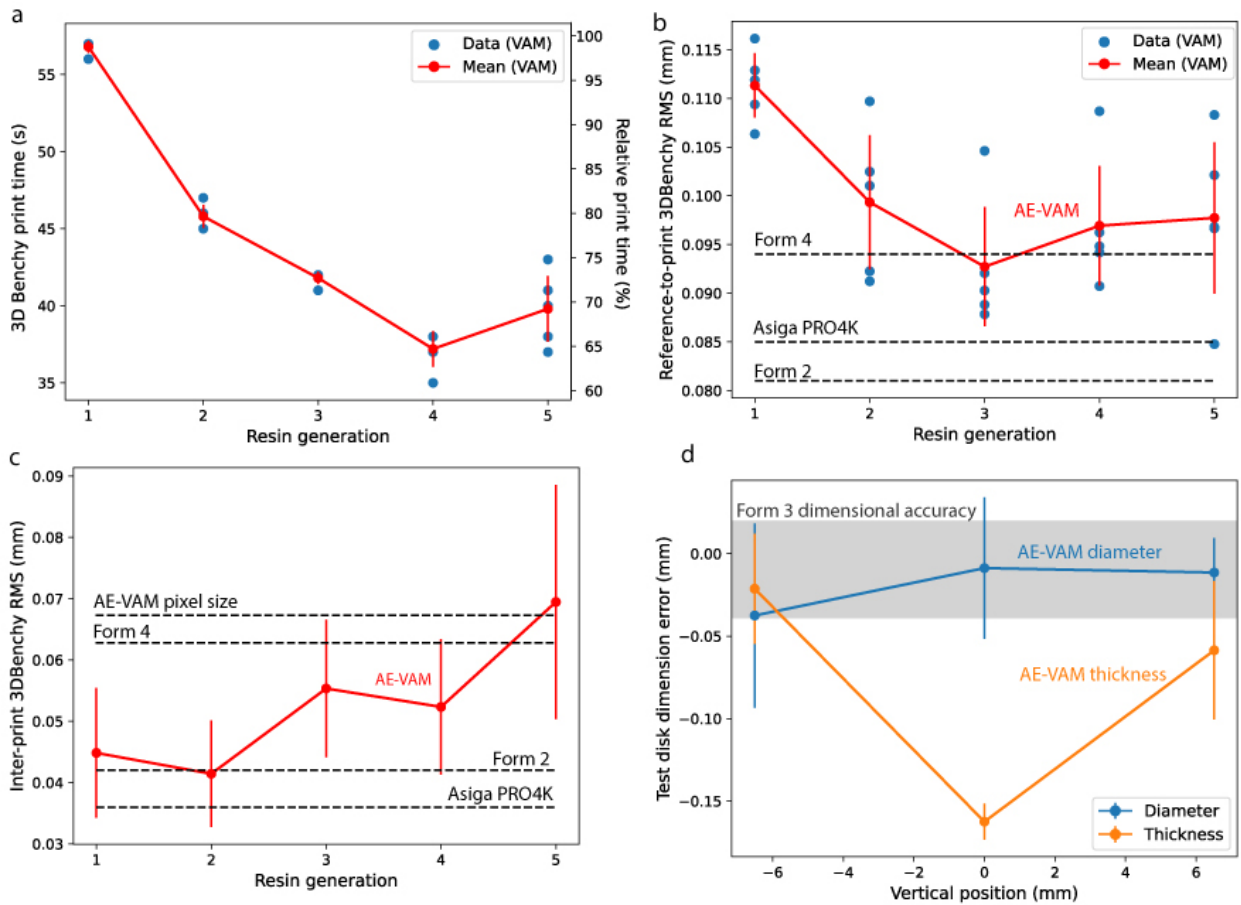

Fig. 4. a) 3DBenchy print time as a function of resin generation. The print time drops significantly as resin is re-used, precluding a fixed exposure time. The blue datapoints show the data for each of the N=5 replicates for each resin generation. The red plot shows the mean and standard deviation (error bars) for each resin generation. b) 3DBenchy reference to print surface RMS deviation as a function of resin generation, as measured by micro x-ray CT. AE-VAM datapoints are shown in blue for all replicates; mean and standard deviations are shown in red. Print accuracy is slightly improved (lower RMS deviation) for re-used resin, likely due to faster prints suffering less from diffusive spread of oxygen. Commercial grade print quality is indicated by the horizontal dashed lines. c) As in (b), but for inter print 3DBenchy RMS deviation, indicating print repeatability. The AE-VAM pixel size used in this work is indicated with the dashed grey line. d) Dimension error of 5mm x 10mm (thickness x diameter) test disks. Error bars represent standard deviations over N=23 total disks (12 gen 1, 11 gen 2). The gray region

represents the reported error range in the Form 3 dimensional accuracy report (Average deviation from ideal ± standard deviation for 4mm and 9mm intended feature size) [28].

**Comparison of AE-VAM to commercial vat photopolymerization 3D printers**

To compare auto-exposure VAM with commercial grade vat-photopolymerization printers, we printed the same size 3DBenchy model on SLA (Formlabs Form 2) and DLP (Formlabs Form 4 and Asiga PRO4K) printers with 0.1mm, 0.05mm and 0.025mm layer thickness, respectively. Resulting x-ray CT scans for the two most accurate printers (Form 2 and Asiga PRO4k) are shown in Fig. 3. The Formlabs printers has manufacturer-calibrated exposure settings, whereas the Asiga PRO 4K requires manual exposure adjustment. We found that over 5 3DBenchy models fabricated on the same build plate in one run (61 mins print time), the Form 2 achieved an average reference-to-print RMS error of 0.081mm, which is slightly more accurate than the 0.100mm achieved with AE-VAM. The inter-print RMS variation of the Form 2 printed 3DBenchy models was 0.042mm compared to 0.045mm and 0.041mm for gen 1 and gen 2 resin AE-VAM prints (Table S1). However, the blind hole (BH) and underside text (UT) do not appear in any 3DBenchy printed with the Form 2, while the through holes in the chimney and rear flagpole holes (CH and FH) are filled, and the chimney and roof are deformed. Results with the Asiga PRO4K DLP (137 mins print time) were similar: 0.085mm reference-to-print RMS error and 0.036mm inter-print RMS variation. Although the Form 4 was the newest and fastest commercial printer tested (51 mins print time), we found that it had the worst print-to-reference RMS (0.094mm) and inter-print RMS (0.062mm) of all commercial printers tested, underscoring the challenge of accurate calibration in 3D printing. For the Asiga PRO4K and Form 4, the underside text and blind hole are missing, and chimney and flagpole holes are partially filled (Fig. 3d, Fig. S3). These printers also struggle to faithfully reproduce the large overhang of the cabin roof. All of these features print correctly with AE-VAM. Dashed horizontal lines in Figs. 4b and c indicate the 3DBenchy reference-to-print and inter-print RMS error, respectively, for all commercial printers tested in this work. These results are summarized in Table S2.

We further tested the repeatability of AE-VAM by printing a series of 5mm × 10mm dimensional test disks. The thickness and diameter of the printed disks were subsequently measured using digital calipers; the resulting dimensional errors are plotted in Fig. 4d. We found a small systematic variation in disk thickness along the vial axis (vertical position): disks in the middle of the write field were thinner compared to the design dimension, and thinner than those at the top and bottom. Further investigation is required to ascertain the cause of this spatial variation, which might be addressed by further projector calibration. Apart from this systematic error, the dimensional accuracy and repeatability of the test disks rivals that obtained by Formlabs' internal testing of the Form 3 3D printer (see gray region in Fig. 4d)[28]. This suggests that if the spatial variation in dose can be further reduced, we can expect auto-exposure VAM to achieve commercial-level repeatability across the write volume.

**Insensitivity to object geometry**

The key to the simplicity of AE-VAM is that by dividing the total scattering signal by the volume of the object in Equation 1, we obtain a *geometry-independent* print state variable. In Figure 5, we show that the exposure threshold value $S_T$ identified in the calibration process is applicable for a wide range of print geometries. Solid models with bulk features such as the Stanford Bunny (Fig. 5a,e), and lattices (Figs. 5b-d, f-h) yield high quality results on the first print attempt – all prints in Fig. 5 were printed first time without any adjustment of exposure time for print refinement (as might be done with many 3D printing technologies), and with generation 5 resin. Touchpoints as small as 0.13mm are achieved in the diamond pentamode lattice[29] shown in Fig. 5d, matching the smallest printable spoke width for $S_T = 0.35$, underscoring the robustness of our approach. To demonstrate the ability to print negative features, we used Flui3d[30] to design simple a microfluidic geometry composed of two embedded channels that cross over each other via a bridge structure. We found that we could repeatably print and clear channels down to ≈ 400 $\mu m$ in width, as shown in Fig. S4. Smaller channel widths could be cleared but not consistently. Further investigation is needed to determine whether this is due to curing of the channel itself, or because more pressure is required to clear the viscous resin from small channels.

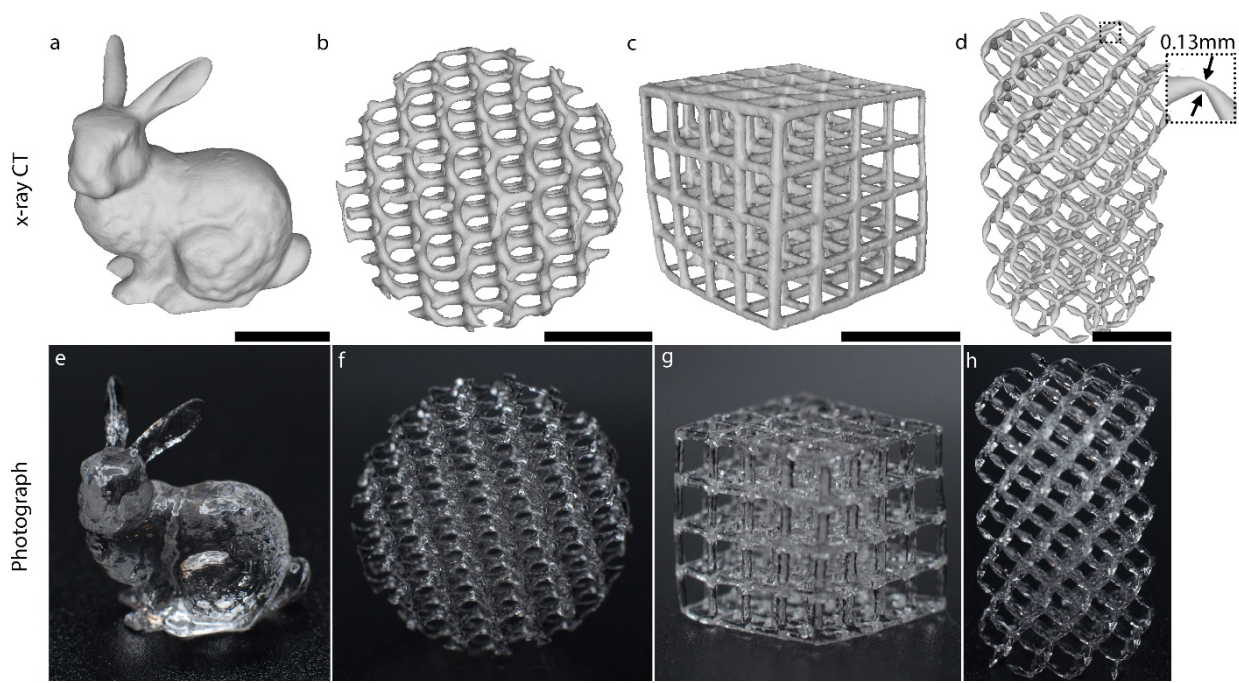

Fig. 5. A variety of auto exposure VAM-printed objects. a) Stanford Bunny. b) Gyroid sphere. c) Cubic lattice. d) Pentamode diamond lattice. Inset: zoom-in of a 0.13mm touch point between struts. e)-h) Photographs of (a)-(d). All scale bars are 5mm.

Commercial 3D printers are often used to fabricate many individual parts that fit together, or mate with an existing part to form a multi-part assembly. With AE-VAM, we achieve the repeatability required to print multiple parts with consistent relative dimensions for assembly. To demonstrate this, we separately printed a ¼-20 cap screw and nut as shown in Fig. 6a. The printed nut can be screwed onto the cap screw (Fig. 6b) and they can also be fastened to standard metal cap screws and nuts (Supplementary Movie 1) with a standard hex key.

We also printed a gear assembly to test the required tolerance to mesh gears. The base of the gear assembly, consisting of two vertical pins, 3mm in diameter, is shown on the left of Fig. 6c. In a separate print, we printed 3 pairs of meshing spur gears. The tolerance, defined as the designed gap between the base pin wall and the inner diameter of the spur gear hole, is shown below for each pair of gears. The backlash of the gears is designed to be equal to the pin tolerance. For the largest tolerance of $100\mu m$, the gears and base can easily be assembled by placing the gears around the pins (Fig. 6d). In this case the gears mesh and spin freely (Supplementary Movie 2), though they do wobble on the pins due to the large tolerance. When the tolerance is reduced to $50\mu m$, the gears can still be assembled on to the base pins and the gears mesh smoothly, but do not spin freely after the rotating force is stopped. For $0\mu m$ tolerance, the gears cannot be assembled onto the pins, as the hole is slightly smaller than the pins. This indicates that to assemble parts, AE-VAM requires a tolerance of approximately $50\mu m$. We note that although this tolerance is below 1 rasterized voxel ($67\mu m$), the voxelization and linear interpolation of the 3D model onto the voxel grid results in modified voxelized representations of the model even if changes to the model are below one voxel in size.

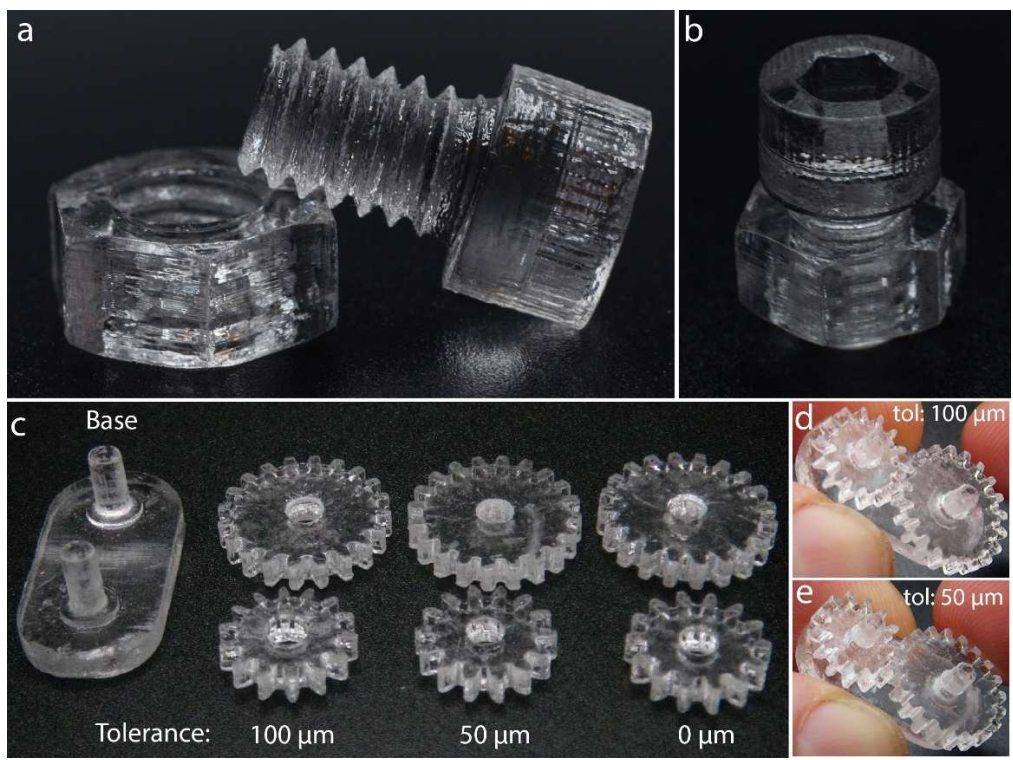

Fig. 6. a) Cap screw and nut printed with auto exposure VAM. b) Fastened cap screw and nut from (a). See Supplementary Movie 1. c) Base and gears printed with auto exposure VAM. Tolerance of through hole and gear spurs are noted below each pair of gears. d) Assembled structure for gears with tolerance $100 \mu m$. See Supplementary Movie 2. e) Assembled structure for gears with tolerance $50 \mu m$. See Supplementary Movie 3.

## Discussion and Conclusion

In tomographic VAM, one must contend with reduced object to background exposure contrast compared to layer-based techniques. This necessitates consistent exposure control for repeatable and accurate results. However, the volumetric nature of light patterning in tomographic VAM results in elevated and unpredictable resin "pre-exposure", effectively reducing the exposure time needed for subsequent prints. Thus, the requirement for tighter exposure control is in opposition with practical reality that the ideal VAM exposure is highly variable. In this work we have solved this crucial challenge, transforming VAM into a repeatable and accurate technique.

We have demonstrated an automatic exposure VAM (AE-VAM) system that can be used without any user intervention for print stopping. To achieve this, we developed a simple, yet highly effective method for measuring print progression enabled by imaging the increase in light scattering intensity over the course of the print. Crucially, AE-VAM requires only the total scattered light signal, making it computationally lightweight and insensitive to the geometry and spatial distribution of the print. This makes AE-VAM extremely robust because it does not require any spatial alignment between reference and projection coordinate systems, nor is it susceptible to artifacts caused by debris in the resin. Moreover, we showed that AE-VAM can accommodate a large change in exposure time when re-using resin: almost 40% when comparing the slowest and fastest 3DBenchy prints (Fig. 4a). This ability to efficiently re-use resin is critical because of the required exposure of out-of-part resin in VAM. AE-VAM allows resin to be re-used until the vial is nearly empty, just like standard vat photopolymerization printers. We presented results for up to generation 5 resin, but we do not track resin beyond generation 5 because the low resin level in a vial requires combining vials that may have different resin generations. We found that there was no significant qualitative print degradation when combining and mixing resin past generation 5 if resin was thoroughly mixed, and the resin level was high enough to accommodate the desired print.

In this work, we made extensive comparisons of AE-VAM to commercial vat photopolymerization printers. We found that AE-VAM has repeatability and accuracy specifications that are within the range measured for commercial systems (see Fig. 4). Furthermore, many small features of the common 3D printer benchmarking object 3DBenchy which are not printable in commercial printers are straightforward to print with AE-VAM (Figs. 3, S3). Printing the 3DBenchy inclined on supports with commercial printers did not render these features printable. Instead, print quality for the Asiga PRO4K and Formlabs Form 4 was degraded due to the presence of support blemishes on the underside of the 3DBenchy (Fig. S5). This underscores another major advantage of VAM: surfaces of all orientations are printable without support structure artifacs.

As expected, AE-VAM significantly outperforms commercial printers in terms of speed. In our benchmarking experiments, the normalized print time per 3DBenchy is approximately 10x faster with AE-VAM (~40-55s) compared to the fastest commercial printer (Form 4, 492s), and is significantly more repeatable but marginally less accurate. When compared to the most repeatable printer (Asiga PRO4K), AE-VAM is 37x faster and correctly prints large overhangs and small voids, but is slightly less accurate overall. We note that light exposure for polymerization takes up only ~11% of the total print time for the Asiga PRO4K (2s/layer + 10s first layer, over 460 layers), compared to 100% for (AE-)VAM. The remaining ~89% of the Asiga PRO4K print time is spent on inter-layer motions such resin reflow under the build plate. Herein lies the fundamental reason for the rapidity of VAM printing.

In this work we focus on implementing and extensively characterizing AE-VAM for one fixed resin formulation. Previous work confirmed that AE-VAM works with similar acrylate resins that show scattering contrast upon curing[31]. More broad testing of the other VAM resins with AE-VAM will be investigated in the future.

This work sets a baseline repeatability and accuracy specification for (AE-)VAM systems which can be compared to commercial vat photopolymerization systems, a significant milestone in maturing tomographic VAM to a commercial level.

## Methods

### Resin

The DUDMA/PEGDA resin formulation consisted of 80 wt% diurethane dimethacrylate (DUDMA, Esstech Inc.) and 20 wt% poly(ethylene glycol) diacrylate (Mn = 700 g/mol, PEGDA 700, Sigma-Aldrich), with 1.75 mM ethyl (2,4,6-trimethylbenzoyl) phenylphosphinate (TPO-L, Oakwood Chemical) as the photoinitiator. Refractive index measurements were taken using a Schmidt-Haensch ATR-BR refractometer, yielding an RI of 1.49962 at 405 nm. The viscosity of the resin was measured to be 1652 cP under standard temperature and pressure (STP) using a TA Instruments Discovery HR20 Rheometer.

Three batches of resin were mixed for the prints in this paper. The calibration prints in Fig. 2 were printed in one batch, the 3DBenchys with another batch, and the prints in Figs. 5 and 6 were printed with a third batch.

### AE-VAM projection calculation and printing

Prints were performed in nominally 1-inch cylindrical borosilicate vials. The actual measured outer diameter of the vials was 24.8 mm. The vial is affixed to the rotation stage with a custom 3D printed vial holder. There is no index matching bath used in this printer.

Projection calculation proceeds as follows. We first rasterize an STL file describing the object using the python package Trimesh[32], resulting in a 3D Numpy matrix (the target geometry). A non-zero background is added to this target geometry. This enables pseudo-negative projected dose (dose that is below the non—zero background) which improves print fidelity. For all prints in this paper, the background dose was 60% of the object dose. This target geometry is pre-compensated for diffusive effects using a previously reported technique[9]. 2D slices along the vertical axis of the object are then radon-transformed using scikit-image[33] with 1 degree sampling. Each Radon-transformed slice is then subjected to a ramp filter, yielding filtered projections that contain negative values which are clipped at 0. The estimated dose is then simulated by numerically backprojecting these nonnegative filtered projections using the iradon function in scikit-image. We then normalize the deconvolved target geometry by this simulated dose – applying a multiplicative gain in object space that increases the desired dose in regions where the simulated dose undershoots and vice versa. New projections are then calculated using this normalized and deconvolved target geometry in the same manner as above. The resulting projections are subsequently corrected for vial distortion and then recalled by the projector during printing[23].

To print, the previously calculated projections are projected through the rotating vial. The rotation rate of the vial was 40 degrees/second for all prints in this paper. The side scatter from the red LED (Thorlabs M660L4) is monitored in real time with a camera (FLIR GS3-U3-23S6M) as detailed in the main text. The LED is approximately collimated with a 30mm focal length plano convex lens. LED power at the bottom of the vial was measured to be 95 $mW/cm^2$. Once the side scatter level reaches the exposure threshold value, the computer automatically turns off the projector UV illumination via a Python command.

**Washing**

After printing, prints are scoped out of the vial and placed into a bath of 2-propanol for 5 minutes. After, prints are transferred to a fresh bath of 2-propanol and left to sit undisturbed for 20 minutes. Prints are subsequently removed from the bath and left to air dry overnight. The following day they are post-cured in a curing box (AnyCubic) for 60 minutes. At this point, the prints are still tacky to touch. Finally, the parts are submerged into a bath of acetone for 1-2 minutes, and then left to dry. This final acetone bath results in parts that are no longer tacky once dried.

**X-ray CT**

Printed samples were scanned with a Skyscan 1275 micro-computed tomography system (Bruker) at a pixel resolution of 25 µm. Image reconstruction was performed using Skyscan NRecon software (Bruker), and signed-distance fields were extracted from the resulting scans using MeshLab.

**Commercial printer printing**

**Formlabs printers**

3DBenchy models, each measuring 11.5 mm in height and without support structures, were printed using Formlabs Form 2 and Form 4 printers. Five samples were printed on the same build plate in one run with the Form 2 using Formlabs Clear Resin V4. A layer thickness of 100 µm and default printing parameters. Similarly, five samples were printed in one run with the Form 4 with Formlabs Clear Resin V5 using a layer thickness of 50 µm and default settings. The layer thicknesses were selected based on optimal print quality determined after testing thicknesses of 25, 50, and 100 µm. Following printing, all samples were washed in isopropyl alcohol (IPA) for 10 minutes and post-cured in a Formlabs Form Cure chamber at 60 °C for 15 minutes, adhering to the manufacturer's recommendations.

**Asiga printer**

The same 3DBenchy models were printed using an Asiga Pro 4K UV385 DLP printer. All 5 models were printed on the bed (no support structures) in one run. The optimized printing parameters included a light intensity of 17 mW/cm², a slice thickness of 25 µm, an exposure time of 2 seconds per layer, a burn-in exposure time of 10 seconds for the initial layer, and one burn-in layer. These settings were determined to produce the highest print quality after iterative testing and adjustment. During printing, Formlabs Clear Resin V4 was preheated to 25 °C.

Post-processing involved washing the samples in IPA for 10 minutes and curing them in a Formlabs Form Cure chamber at 60 °C for 15 minutes, in line with the manufacturer's guidelines.

3DBenchys with support structures were printed with the Formlabs Form 4 and Asiga PRO4K printers, as shown in Fig. S5. These prints were not used to generate any of the data reported elsewhere in the paper.

**Funding.**  This work was funded by the National Research Council Ideation program.

**Conflict of Interest**.  Several authors are inventors of provisional patents related to this work.  The authors declare no other conflicts interest.

**Data availability**.  The data presented in this paper are available upon reasonable request.

**Supplementary Information**

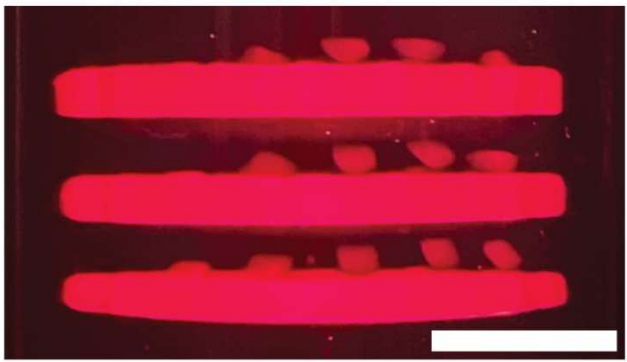

Fig. S1. a) Optical scattering photograph of a calibration disk approximately 2 minutes after exposure is terminated. Gapped disks float above the main disk due to different sedimentation rates of the disconnected bodies. Scale bar is 5mm.

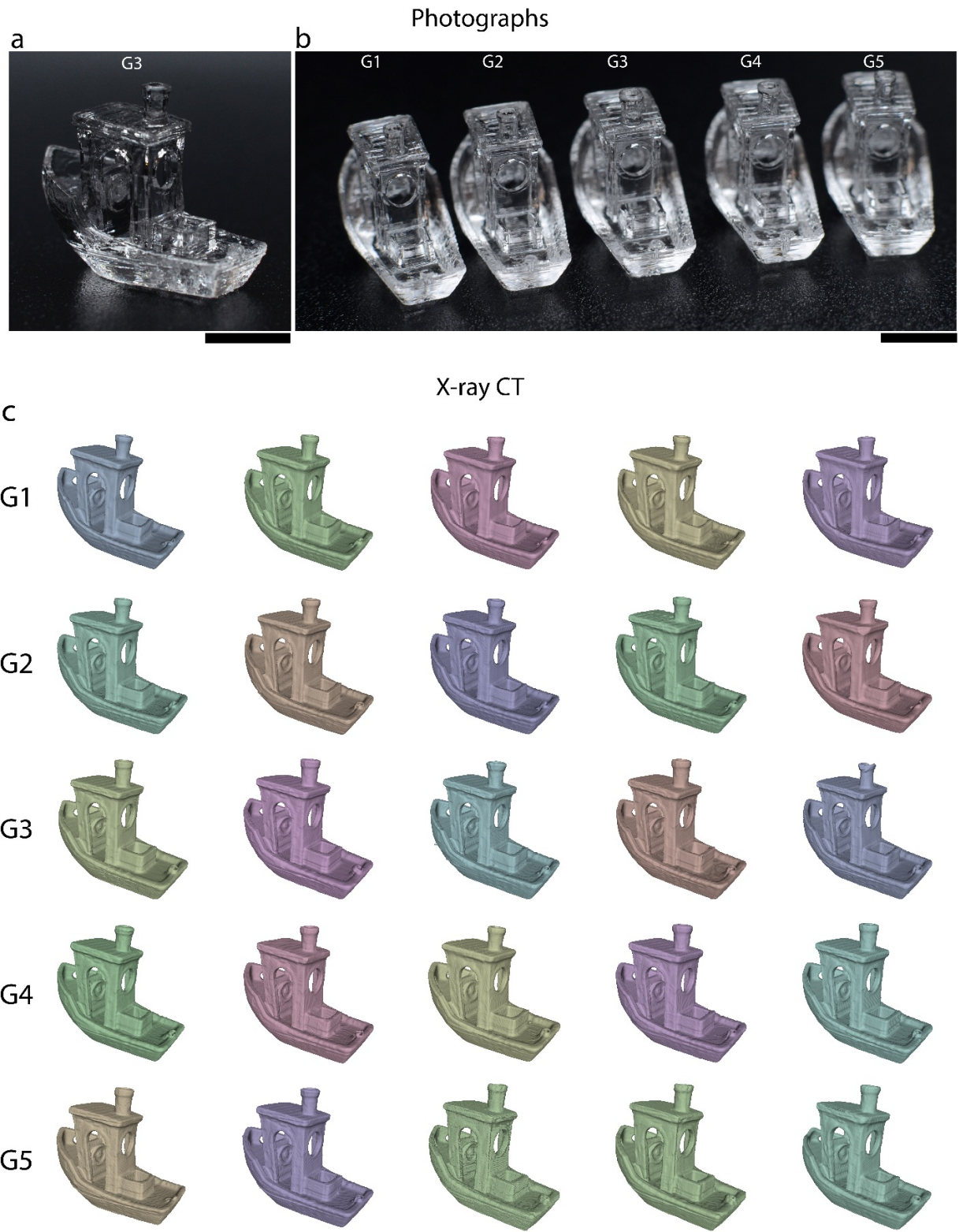

Fig. S2. a) Photographs of a typical AE-VAM printed 3DBenchy using gen 3 (G3) resin. Scale bar is 5 mm. b) Photograph of selected AE-VAM printed 3DBenchys for gen 1 (G1) – gen 5 (G5) resin. Scale bar is 5 mm. c) Micro x-ray CT renderings of auto-

exposure VAM printed 3DBenchy models. The row label refers to the generation of resin, with G1 being fresh resin, G5 being 4x re-used resin.

## Form 4 3DBenchy

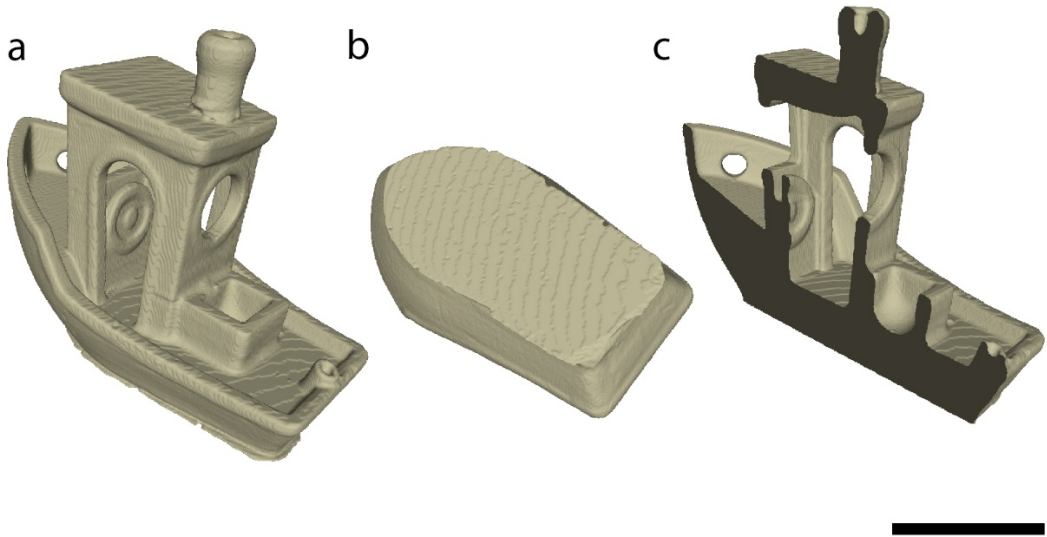

Fig. S3. a) A 3DBenchy printed with a Formlabs Form 4 3D printer. Note the deformation of the chimney in comparison to the reference model, VAM print and Asiga PRO4K print in Figs. 3a,b,d. b) Underside of the same print in (a), showing the lack of the blind hole and underside lettering that is in the reference geometry and VAM print (Fig. 3). c) Cross-section of the Form 4 3DBenchy. Note the rear flagpole holder is filled it, the rear blind hole is not present, and the chimney is filled. Scale bar is 5mm.

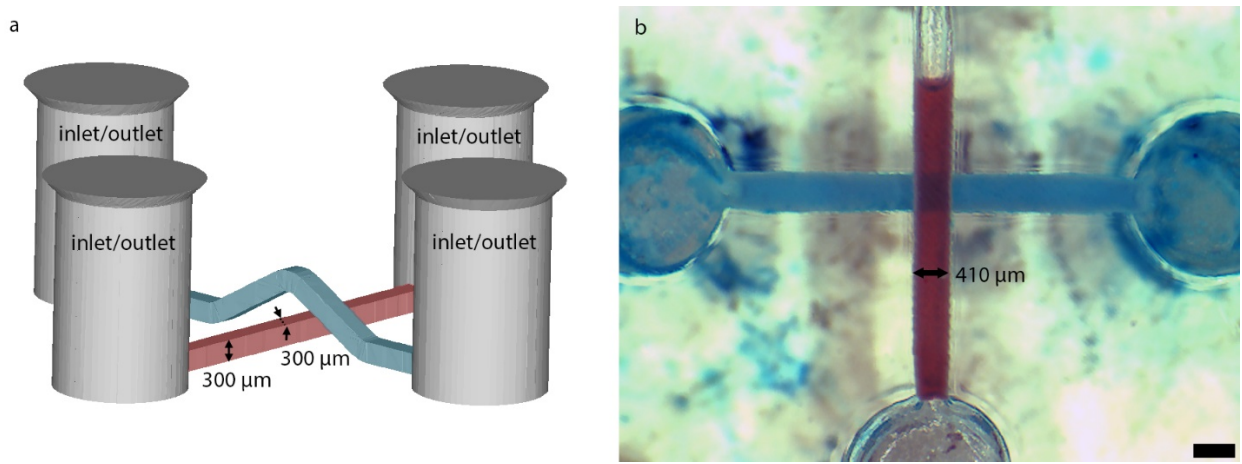

Fig. S4. a) 3D rendering of channel bridge design, comprised of an upper (blue) and lower (red) channel, each with a $300\ \mu m$ square cross section. The upper channel crosses over the lower channel with a bridge structure. b) Optical micrograph of the VAM-printed channel structure in (a). The upper and lower channels are filled with blue and red channel, respectively. The blue channel appears to cross behind the red channel due to the bridge structure visible in (a). The printed width of the red channel ($410\ \mu m$) is indicated. This is slightly larger than the design width $300 \mu m$, and matches with the smallest gap that can be consistently achieved in the calibration geometry (Fig. 2b). Scale bar is $500\ \mu m$.

3D Benchy prints with supports on commercial printers

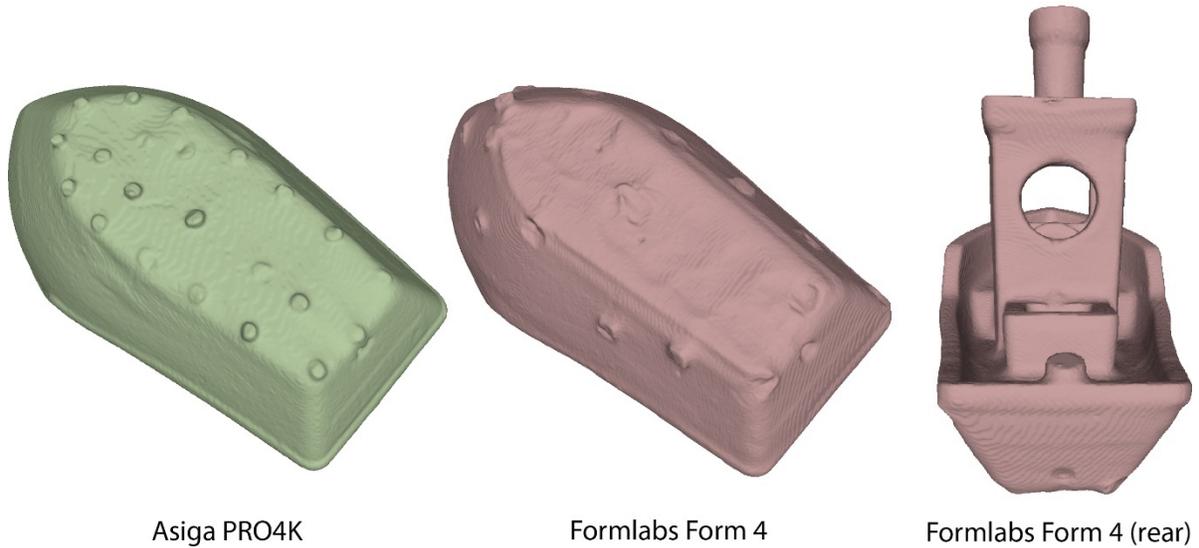

Asiga PRO4K      Formlabs Form 4      Formlabs Form 4 (rear)

Fig. S5. X-ray CT renderings of the bottom of 3DBenchy prints using supports with commercial printers. Significant support structure artifacts remain after washing and underside text is not present. The rear of the Form 4 print is displayed to show that the blind hole and flagpole holder remain filled, even when printing on supports at manufacturer's software optimized print orientation.

## 1. AE-VAM 3DBenchy print details

|  | Mean reference-to-print RMS ($\pm$std) [mm] | Mean reference-to-print deviation [mm] | Mean inter-print RMS ($\pm$std) [mm] | Mean print time ($\pm$std) [s] |
|---|---|---|---|---|
| Generation 1 resin (fresh) | 0.111 ($\pm$0.004) | -0.028 | 0.045 ($\pm$0.011) | 56.8 ($\pm$ 0.4) |
| Generation 2 resin | 0.099 ($\pm$0.008) | -0.016 | 0.041 ($\pm$0.009) | 45.8 ($\pm$ 0.8) |
| Generation 3 resin | 0.093 ($\pm$0.007) | -0.010 | 0.055 ($\pm$0.011) | 41.8 ($\pm$ 0.4) |
| Generation 4 resin | 0.097 ($\pm$0.007) | -0.005 | 0.052 ($\pm$0.011) | 37.2 ($\pm$ 1.3) |
| Generation 5 resin | 0.098 ($\pm$0.009) | -0.004 | 0.069 ($\pm$0.020) | 39.8 ($\pm$ 2.4) |
| ***All generations*** | **0.100 ($\pm 0.009$)** | **-0.013** | **0.053 ($\pm 0.015$)** | **44.3 ($\pm$ 7.1)** |

Table S1. 3DBenchy print accuracy as quantified by the RMS deviation between reference file and print geometry, and inter-print RMS deviation, respectively. Mean values and standard deviations are calculated over N=5 prints for each generation of resin.

## 2. 3DBenchy print summary

|  | Form 2 | Form 4 | Asiga PRO4K | AE-VAM (G2) |
|---|---|---|---|---|
| Slicing thickness [μm] | 100 | 25 | 50 | 67 |
| Reference-to-print RMS ($\pm$std) [mm] | **0.081** ($\pm$0.004) | 0.094 ($\pm$0.002) | 0.085 ($\pm$0.004) | 0.099 ($\pm$0.008) |
| Inter-print RMS ($\pm$std) [mm] | 0.042 ($\pm$0.008) | 0.062 ($\pm$0.022) | **0.036** ($\pm$0.008) | 0.041 ($\pm$0.009) |
| Print time / 3DBenchy [s] | 732 | 492 | 1644 | **46** |
| Chimney | Partially filled, distorted. | Partially filled, distorted on some prints. | Partially filled | **Clear** |
| Underside Text | None | None | None | **Visible** |
| Blind Hole | None | None | None | **Visible** |
| Flagpole Holder | Filled | Filled | Partially filled | **Clear, present in 19/25 prints.** |
| Cabin roof overhang | Bowed | Bowed | Bowed | **Flat** |

Table S2. Summary of dimensional accuracy, repeatability, and speed of 3DBenchy prints. Qualitative appearance of small subtle features is also noted. The printer with the best result in each category is bolded.

3. **Dimensional test disks**

| Vertical Position [mm] | Mean thickness error [mm] | Thickness std [mm] | Mean diameter error [mm] | Diameter std [mm] |
|---|---|---|---|---|
| +6.2 | -0.06 | 0.04 | -0.01 | 0.02 |
| 0 | -0.16 | 0.01 | -0.01 | 0.04 |
| -6.2 | -0.02 | 0.03 | -0.04 | 0.06 |
| All | -0.08 | 0.07 | -0.02 | 0.05 |

Table S3. Mean measured thickness and diameter of all G1 (N=12) and G2 (N=11) dimensional test disks printed with AE-VAM.